\documentclass[aip,apl,reprint]{revtex4-1}

\usepackage{graphicx}
\usepackage{dcolumn}
\usepackage{bm}

\usepackage[utf8]{inputenc}
\usepackage[T1]{fontenc}
\usepackage{mathptmx}
\usepackage{etoolbox}
\usepackage{hyperref}

\makeatletter
\def\@email#1#2{%
 \endgroup
 \patchcmd{\titleblock@produce}
  {\frontmatter@RRAPformat}
  {\frontmatter@RRAPformat{\produce@RRAP{*#1\href{mailto:#2}{#2}}}\frontmatter@RRAPformat}
  {}{}
}%
\makeatother
\begin{document}


\title[]{Temperature-dependent acoustic loss at microwave frequencies in thin-film lithium niobate}
\author{Qixuan Lin}
 \affiliation{Department of Electrical and Computer Engineering, University of Washington, Seattle, WA 98115, USA}
 
 \author{Yue Yu}
 \affiliation{Department of Electrical and Computer Engineering, University of Washington, Seattle, WA 98115, USA}
 
 \author{Alejandra Guedeja-Marr\'on}
 \affiliation{ Materials Sciences and Engineering Department, University of Washington, Seattle, WA 98115, USA.}
 
 \author{Catalina Scolnic}
 \affiliation{ Materials Sciences and Engineering Department, University of Washington, Seattle, WA 98115, USA.}
 
 \author{Haoqin Deng}
 \affiliation{Department of Electrical and Computer Engineering, University of Washington, Seattle, WA 98115, USA}
 
 \author{Shucheng Fang}
 \affiliation{Department of Electrical and Computer Engineering, University of Washington, Seattle, WA 98115, USA}
 
 \author{Yibing Zhou}
 \affiliation{Department of Electrical and Computer Engineering, University of Washington, Seattle, WA 98115, USA}
 
 \author{Bingzhao Li}
 \affiliation{Department of Electrical and Computer Engineering, University of Washington, Seattle, WA 98115, USA}

 \author{Juan Carlos Idrobo}
 \affiliation{ Materials Sciences and Engineering Department, University of Washington, Seattle, WA 98115, USA.}
 
 \affiliation{Physical and Computational Sciences Directorate, Pacific Northwest National Laboratory, Richland, WA 99354, USA.}
 
\author{Mo Li}%
\email{moli96@uw.edu}
\affiliation{Department of Electrical and Computer Engineering, University of Washington, Seattle, WA 98115, USA
}%
\affiliation{Department of Physics, University of Washington, Seattle, WA 98115, USA
}%

\date{\today}

\begin{abstract}
Thin-film lithium niobate (TFLN) has emerged as a versatile platform for phononic and photonic devices with applications ranging from classical signal processing to quantum technologies. However, acoustic loss fundamentally limits the performance of acoustic devices on TFLN platforms, yet its physical origin remains insufficiently understood. Here, we systematically investigate acoustic propagation loss in various TFLN platforms, including lithium niobate on insulator (LNOI), lithium niobate on sapphire (LNOS), suspended LN thin films, and bulk LN at gigahertz frequencies over temperatures ranging from 4 K to above room temperature. Using a delay-line method, we extract frequency- and temperature-dependent losses for Rayleigh, shear-horizontal, and Lamb modes. We observe an anomalous non-monotonic temperature dependence in LNOI that closely resembles acoustic loss in amorphous materials, indicating a dominant loss channel associated with the buried oxide layer at low temperatures. At elevated temperatures, the loss converges to the Akhiezer damping governed by phonon-phonon interactions. High-resolution electron microscopy further reveals nanoscale interfacial crystal impurities that may contribute to the increased acoustic loss in TFLN platforms relative to bulk LN. These results elucidate the acoustic loss mechanisms in TFLN and provide guidelines for designing low-loss acoustic devices.

\end{abstract}


\maketitle 
Surface acoustic waves (SAWs), including Rayleigh, Lamb, Love, and shear-horizontal (SH) waves, have become increasingly important for a wide range of modern micro- and nano-mechanical technologies. Owing to their velocities being approximately five orders of magnitude lower than those of electromagnetic waves, acoustic waves can be excited at microwave frequencies with micrometer-scale wavelengths. This unique property enables compact SAW microwave filters \cite{lu2021rf,tong20246}, acousto-optic devices \cite{li2023frequency,ye2025integrated}, and quantum devices \cite{kurizki2015quantum,satzinger2018quantum,bienfait2019phonon,qiao2023splitting,bozkurt2023quantum}. In addition, SAWs generate oscillating strain, displacement, and, in piezoelectric materials, accompanying electric fields, traveling in a continuous medium. These fields can strongly interact with particles \cite{lange2008surface}, solid-state quantum systems, such as excitons, electron spins, and defect-related states \cite{aspelmeyer2014cavity,golter2016coupling,maity2020coherent,sasaki2021magnetization,peng2022long,chen2024phonon}. SAW also couples with light, especially guided modes in integrated photonics, thereby enabling optomechanical and acousto-optic devices for advanced optical sensing \cite{tang2023single,li2023frequency}, communication \cite{liu2019electromechanical,lin2025optical,ye2025integrated,selvin2025acoustic}, and quantum transduction \cite{kurizki2015quantum,jiang2020efficient,barzanjeh2022optomechanics}.
 
Among many piezoelectric materials that are employed for SAW devices, lithium niobate (LN) stands out for its strong piezoelectricity and excellent mechanical and optical properties \cite{slobodnik1970microwave,lu2021rf,shao2022electrical,qiao2023splitting}. SAWs with low propagation loss can be efficiently generated on LN, and high-performance SAW filters have been demonstrated on bulk LN. Recently, thin-film lithium niobate (TFLN) platforms, including lithium niobate on insulator (LNOI), lithium niobate on sapphire (LNOS), and suspended LN thin film, have emerged for many acoustic and optical applications. They support co-guided optical and acoustic modes with enhanced mode confinements and interaction strength. They enable efficient acousto-optic \cite{li2023frequency,ye2025integrated} and piezo-optomechanical devices \cite{shen2020high,jiang2020efficient}. In addition, non-suspended TFLN substrates offer improved mechanical robustness and power-handling capability, making them particularly attractive for integrated SAW-photonic platforms.

However, TFLN platforms, especially non-suspended ones, suffer from a significantly higher acoustic loss than that of bulk LN \cite{lu2021gigahertz,luschmann2023surface,lin2025optical}. The resulting loss of acoustic energy degrades device performance, for example, reducing acousto-optic interaction efficiency \cite{lin2025optical}. Acoustic loss mechanisms in bulk lithium niobate (LN) have been extensively studied. For instance, Slobodnik et al. \cite{slobodnik1970microwave} reported that, for Rayleigh-mode surface acoustic waves in bulk LN at 1 GHz, the total propagation loss consists of a temperature-dependent propagation loss of 0.23~dB/mm at room temperature, an air-loading loss of 0.016~dB/mm, and a temperature-independent intrinsic propagation loss of 0.016~dB/mm. In contrast, a comprehensive study of the underlying acoustic loss mechanisms in TFLN platforms has not yet been reported and is therefore imperative.

\begin{figure*}[t]
\centering
\includegraphics[width=5.in]{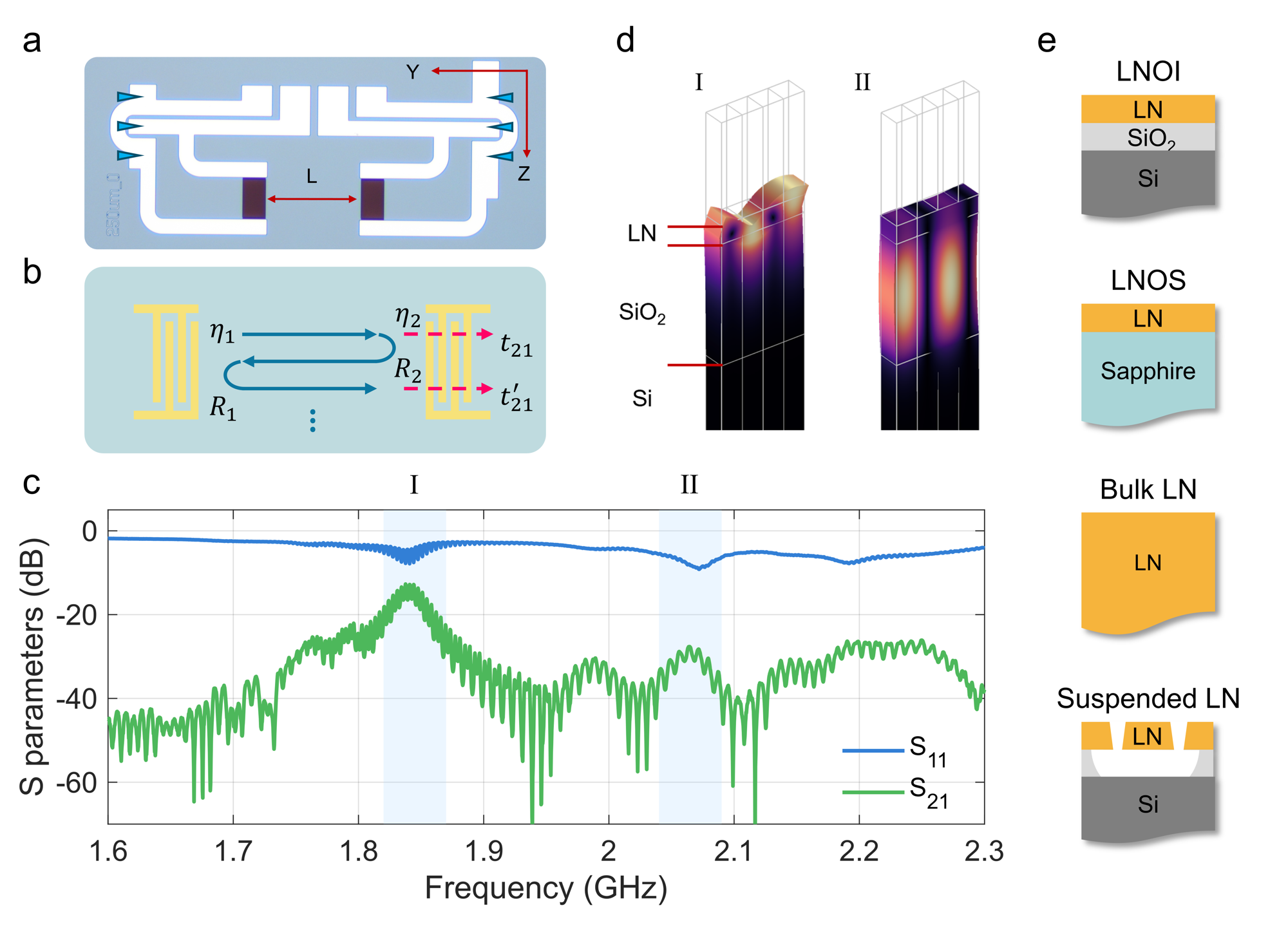}
\caption{\textbf{Device design and measurement principles.} (a). An optical microscope image of an IDT pair fabricated on X-cut LNOI substrate. Two slightly chirped IDTs are fabricated along the LN $y$-axis with a delay length $L$. The blue triangles indicate the probing locations using a pair of GSG probes. (b) Schematic illustration of the acoustic propagation loss measurement scheme using a delay line. (c). Representative S-parameter measurement results of the device. Regions I and II correspond to the excitation of Rayleigh and shear-horizontal acoustic modes, respectively. (d). Finite-element-method (FEM) simulations showing the displacement field distributions of the two acoustic modes. (e) Schematic illustrations of the material stacks of the samples studied in this work: (A) LNOI, (B) LNOS, (C) bulk LN, and (D) free-standing LN thin film.}
\label{Fig1}
\end{figure*}

Acoustic loss in solids arises from various mechanisms, such as phonon-phonon interaction, air loading, and scattering by material impurities \cite{slobodnik1970microwave,biryukov1995surface}. At low temperatures, acoustic loss due to coupling to two-level systems (TLS) becomes significant \cite{manenti2016surface,wollack2021loss,luschmann2023surface,chen2024phonon}. Distinguishing among these loss mechanisms is essential for the design and optimization of the mechanical performance of acoustic devices on TFLN substrates.

In this work, we systematically analyze acoustic loss mechanisms of TFLN by measuring propagation loss in four different platforms: (A) X-cut LNOI consisting of a 300-nm-thick LN layer on a 2-$\mu$m buried oxide on a silicon handle wafer, a structure which is widely used for integrated photonics; (B) X-cut LNOS consisting of a 400-nm-thick LN layer on sapphire handle wafer; (C) bulk 128$^\circ$ YX-cut LN; and (D) a free-standing LN thin film realized by undercutting the LNOI structure in platform A. Schematic illustrations of the material stacks of the samples are shown in Fig.~\ref{Fig1}e. By characterizing both the frequency and temperature dependence of the acoustic propagation loss across these material platforms, we gain insight into the acoustic loss mechanisms in TFLN devices.

The acoustic propagation loss is measured using a delay-line method. A pair of interdigital transducers (IDTs) separated by a delay length $L$ is fabricated to excite and receive acoustic waves, as shown in Fig. \ref{Fig1}(a) (see Supplementary Information~\ref{smsec:fab} for the fabrication process). On samples A, B, and D, the acoustic waves propagate along the Y-axis of the LN crystal, whereas on sample C, they propagate along the X-axis. A vector network analyzer (VNA) is used to excite one IDT and measure the response at the other IDT to obtain the device S-parameters (see Supplementary Information~\ref{smsec:measurement} for the calibration and measurement process). Frequency-domain measurement results can be transformed to the time domain for further analysis.

\begin{figure*}[t]
\centering
\includegraphics[width=6.2in]{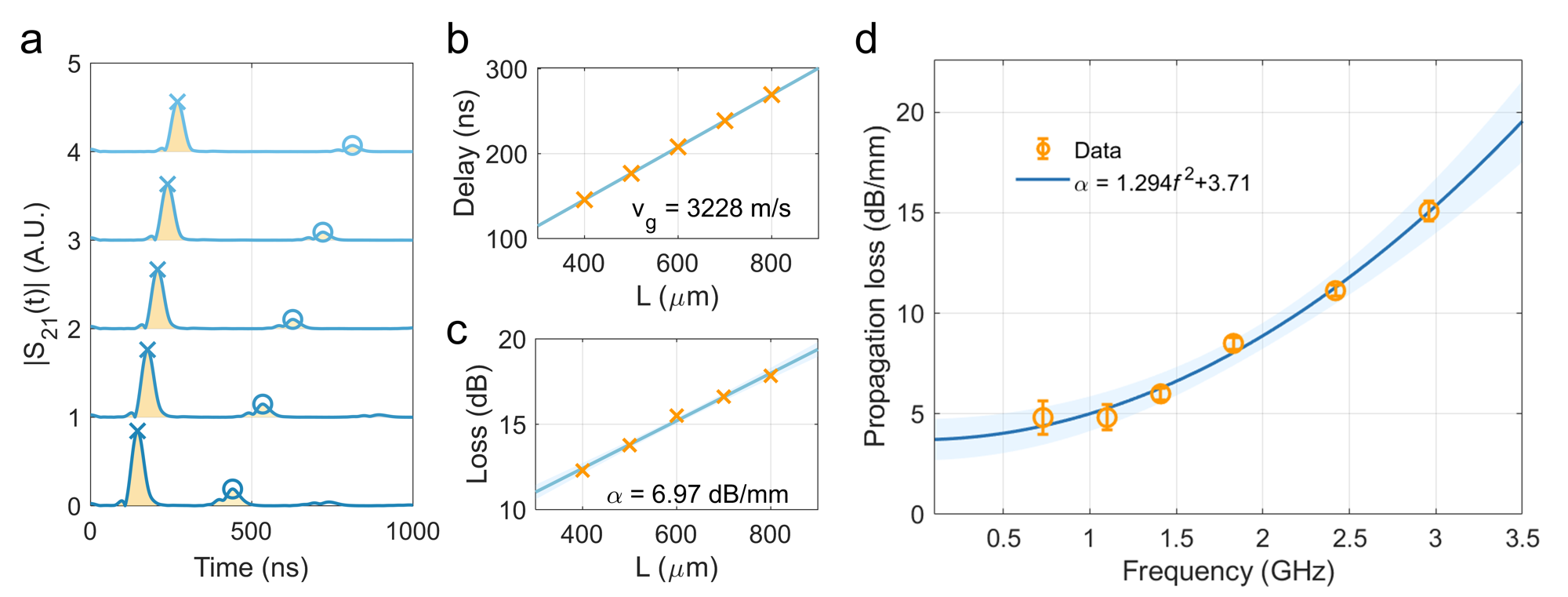}
\caption{\textbf{Acoustic loss measurement using a delay-line method.} (a). Time-domain signals obtained by taking the inverse Fourier transform of the $S_{21}$ response within the 1.82-1.87~GHz frequency range. Each trace corresponds to an IDT pair with a different delay length $L$, varying from $400~\mu \mathrm{m}$ (bottom) to $800~\mu \mathrm{m}$ (top), with vertical offsets applied for clarity. The shaded wave packets indicate acoustic pulses arriving at the receiving IDT via direct transmission or multiple round-trip reflections. Cross and circle markers denote the peak amplitudes of the first and the second arrival wave packets, respectively. Measurements are performed at room temperature. (b). Acoustic group velocity extracted from a linear fit of the delay time of the first wave packet as a function of the corresponding delay length. (c). A 6.97 dB/mm propagation loss extracted from a linear fit of the measured amplitude ratio between the first and the second wave packets at different delay lengths $L$. (d). Frequency dependence of acoustic propagation loss, fitted with a quadratic function of frequency. The shaded regions in this figure and the following indicate the 95\% confidence interval, unless otherwise indicated.}
\label{Fig2}
\end{figure*}

Fig. \ref{Fig1}(b) illustrates the delay line measurement principle for acoustic propagation loss. An acoustic pulse generated by the first IDT travels in the material and arrives at the second IDT after a transit time $\tau=L/v_g$, where $L$ is the distance between the two IDTs and $v_g$ is the acoustic group velocity. At the receiving IDT, part of the acoustic energy is detected while part of it is reflected, which travels back and forth to form a sequence of echoes with sequential delay times of $2\tau$. The transmission coefficient of the electrical signal through the first-arriving acoustic wave packet can be written as
\begin{equation}
    t_{21}=\eta_1\eta_2e^{-\alpha L}
    \label{eq:first_arrive}
\end{equation}
and the transmission coefficient of the second arriving wave packet, after one reflection at each IDT, is
\begin{equation}
    t'_{21}=\eta_1\eta_2r_1r_2e^{-3\alpha L}
    \label{eq:second_arrive}
\end{equation}
where $\eta_1$ ($\eta_2$) denotes the acoustic wave excitation (receiving) efficiencies by the excitation (receiving) IDT, and $r_1$ and $r_2$ are reflectivities of acoustic waves at the two IDTs, respectively. Therefore, the ratio of the amplitudes of two wave packets is:
\begin{equation}
    \gamma(L)=\frac{t_{21}}{t'_{21}}=e^{2\alpha L}/(r_1r_2)
    \label{eq:echo_ratio}
\end{equation}
, where $\eta_1$ ($\eta_2$), which will need independent calibration, are canceled out. We fabricate a set of devices with identical IDT designs and fabrication processes, but with varying delay lengths $L$. Assuming that the acoustic reflectivities $r_1$ and $r_2$ remain the same across these devices, the acoustic propagation loss $\alpha$ is extracted by fitting $\gamma(L)$ as a function of $L$. It should be noted that the acoustic loss is extracted from free-propagating delay-line measurements without lateral confinement. As a result, the reported loss includes not only intrinsic propagation loss but also additional contributions arising from the measurement configuration, associated with beam walk-off and divergence, which reduce the overlap between the acoustic field and the receiving transducer. A detailed evaluation of these effects is provided in Supplementary Information~\ref{smsec:walkoff}.

Fig.~\ref{Fig1}(c) shows the measured S-parameters, reflection coefficient $S_{11}$, and transmission coefficient $S_{21}$, of a representative device fabricated on sample A. Each IDT consists of 25 pairs of electrode fingers with a slightly chirped pitch ranging from 1.5~$\mu$m to 1.65~$\mu$m and a constant duty cycle of 50\%. The two shaded regions, I and II, correspond to the excitation of the fundamental Rayleigh mode and shear-horizontal (SH) mode, respectively. The corresponding displacement field distributions for the two modes, simulated using the finite-element method, are shown in Fig.~\ref{Fig1}(d).

We first focus on the Rayleigh mode in the 1.82–1.87~GHz frequency range. All measurements are conducted at room temperature unless otherwise noted. By applying a window over the selected frequency band and performing an inverse Fourier transform of the measured $S_{21}(f)$, we obtain the time-domain response of the Rayleigh mode, as shown in Fig.~\ref{Fig2}(a). The two peaks in each time-domain response correspond to the arrivals of the first and the second acoustic wave packets. Fitting the delay time with the delay length, as shown in Fig. \ref{Fig2}(b), yields a group velocity of $v_g=3,228$~m/s, which agrees well with the simulation. 

We then calculate acoustic loss $\gamma(L)$ for each device. This ratio reflects the accumulated acoustic propagation loss over a propagation distance of $2L$, as plotted in Fig. \ref{Fig2}(c). By performing a linear fit to the measured loss as a function of the propagation length, we extract a Rayleigh-mode propagation loss of 6.97~dB/mm at a central frequency of 1.84~GHz.

\begin{figure*}
\centering
\includegraphics[width=5.9in]{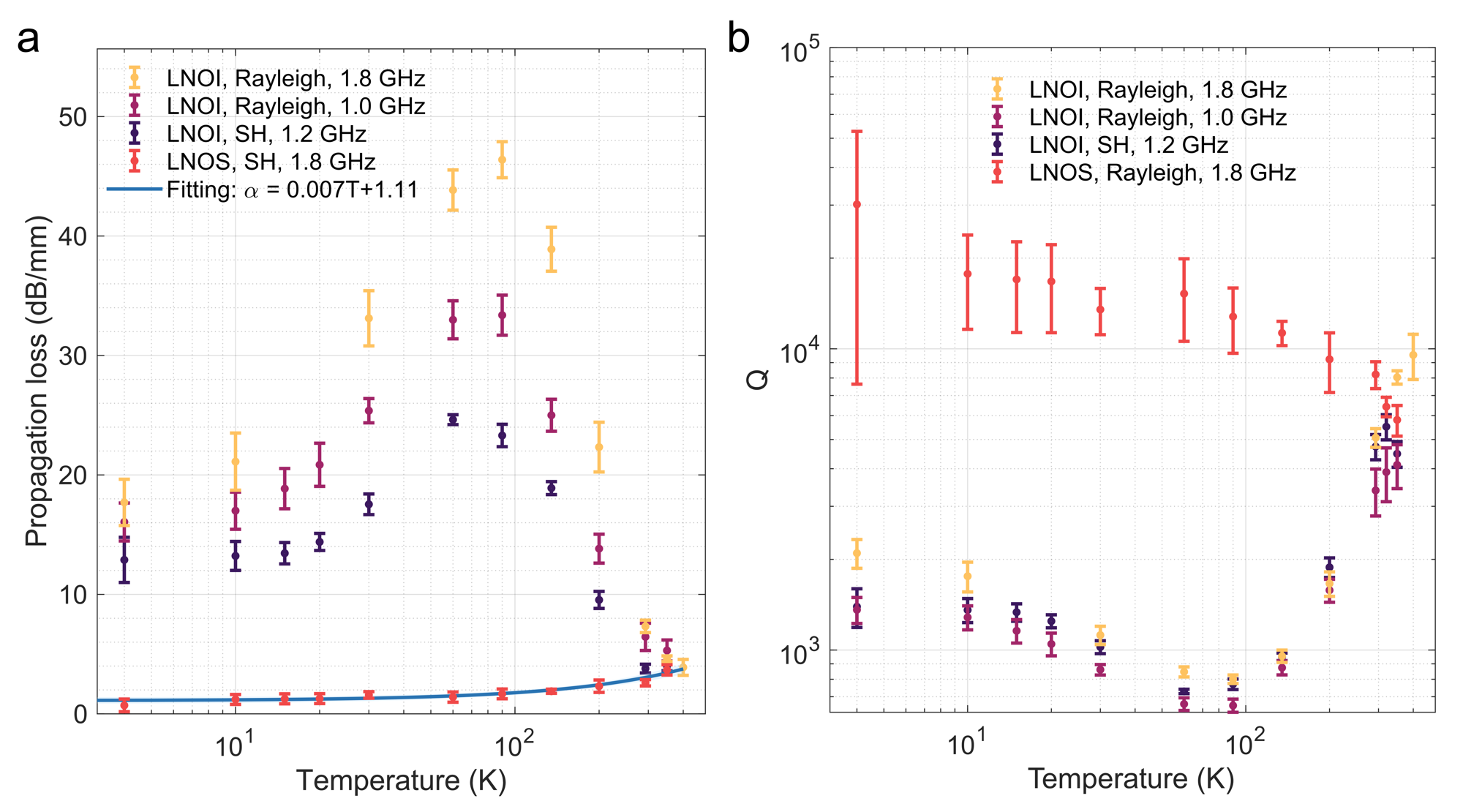}
\caption{\textbf{Temperature dependence of acoustic loss in TFLN platforms.} (a) Acoustic propagation loss of different acoustic modes measured in LNOI and LNOS in the temperature range from 4 to 400 K. (b) The temperature-dependent quality factors extracted from measured acoustic loss in (a), acoustic frequencies, and group velocities of each mode.}
\label{Fig3}
\end{figure*}

With the above measurement scheme, we study the frequency dependence of the acoustic loss. In dielectric crystals, a key intrinsic loss mechanism is due to interaction with thermal phonons in the medium \cite{woodruff1961absorption}. In the case $\omega\tau\ll1$, where $\omega$ is the phonon angular frequency and $\tau$ is the thermal phonon relaxation rate, the acoustic loss is described by the Akhiezer damping model: $\alpha\propto T\omega^2$, where $\alpha$ is the propagation loss in dB/mm. When T decreases, or $\omega$ increases such that $\omega\tau\gg1$, it is in the Landau and Rumer regime~\cite{pomerantz2005ultrasonic}, such that $\alpha\propto T\omega$. In bulk single-domain LN, acoustic loss at room temperature has been reported to follow the Akhiezer loss mechanism with frequency dependence $\alpha\propto \omega^{2}$ up to frequencies approaching tens of gigahertz \cite{bajak1981attenuation,wen1966acoustic}.

To obtain the frequency dependence of acoustic loss, we repeated measurements of the Rayleigh mode at different frequencies on LNOI. As shown in Fig.~\ref{Fig2}(d), the propagation loss reduces quadratically as frequency decreases from 2.96~GHz to 0.73~GHz, with a fitting function of $\alpha_{LNOI,f}=1.294\,f^2+3.71$ in the unit of dB/mm, where $f$ is the acoustic frequency in gigahertz. The frequency-dependent term of $1.294\,f^2$ is consistent with the result measured in crystalline LN, while the frequency-independent residue loss of 3.71~dB/mm indicates that extra loss channels may exist on the LNOI platform.

In addition to frequency dependence, Akhiezer damping is theoretically expected to vary linearly with temperature. In contrast, acoustic loss due to defect scattering can exhibit a strong non-monotonic temperature dependence \cite{pohl2002low}. Defect scattering loss in amorphous materials is attributed to scattering by defect modes with a broad spectral distribution and can be well modeled with the tunneling model \cite{phillips1972tunneling}. The ratio of the phonon wavelength $\lambda$ to the phonon mean free path $l$ of these studied amorphous materials \cite{pohl2002low} falls in the range of $10^{-3}-10^{-2}$ at low temperature.

Motivated by these distinct temperature-dependent loss behaviors, we investigated the acoustic loss of all four platforms as a function of temperature. Because the intrinsic mechanical properties of LN are only weakly dependent on temperature \cite{smith1971temperature}, temperature-dependent measurements are effective to infer the dominant acoustic loss channels in each platform.

\begin{figure*}
\centering
\includegraphics[width=5.9in]{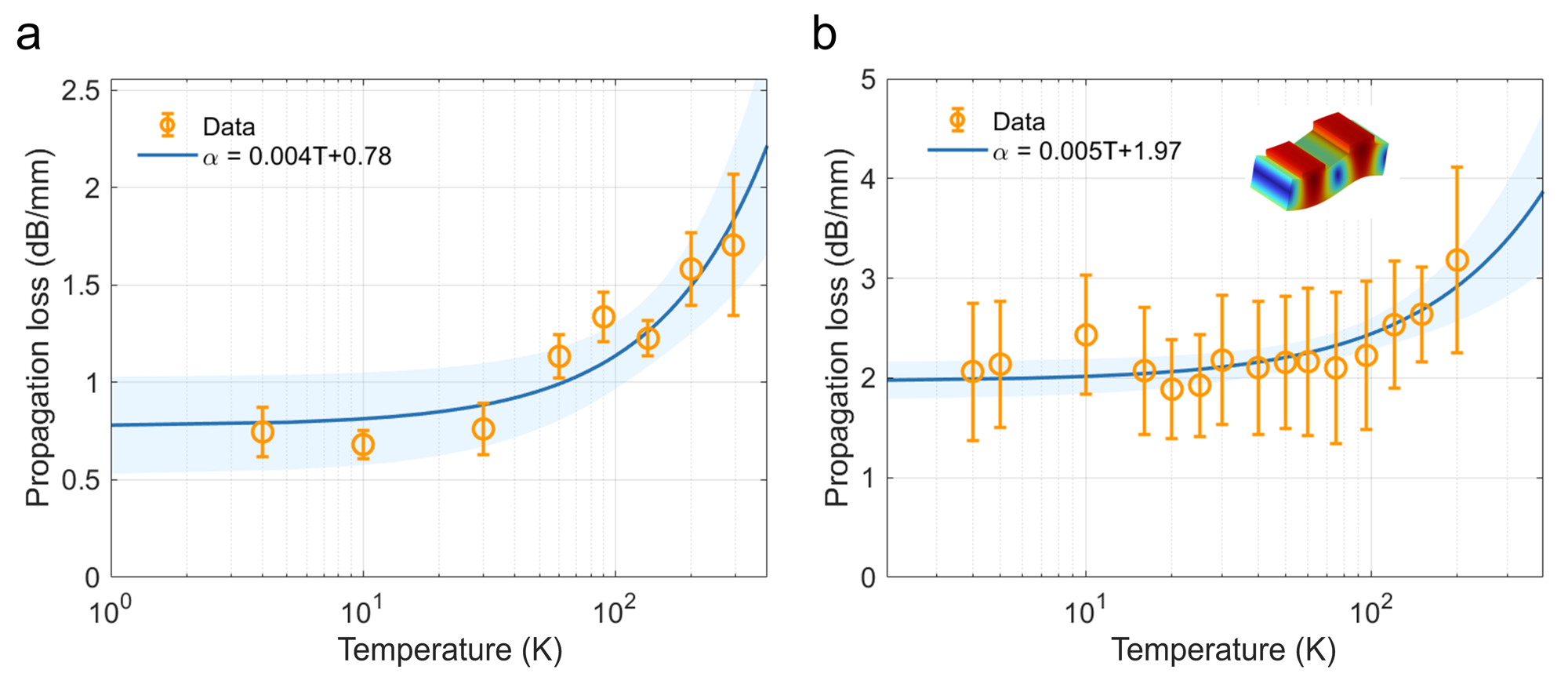}
\caption{Temperature-dependent acoustic propagation loss of (a). Fundamental Rayleigh mode of the bulk LN and (b). Fundamental asymmetric Lamb (A0) mode of the suspended LN. The inset shows the FEM simulation of the A0 mode profile of a 300 nm thick LN with 100 nm Au electrodes on top.}
\label{Fig4}
\end{figure*}
Fig. \ref{Fig3}(a) shows the temperature dependence of acoustic loss in LNOI and LNOS measured from 4 K to 400 K. In LNOS, the acoustic loss decreases monotonically with decreasing temperature. The acoustic loss in LNOS exhibits an approximately linear dependence on temperature, consistent with Akhiezer damping. In contrast, the acoustic loss in LNOI exhibits a non-monotonic temperature dependence. For both Rayleigh and SH modes, the acoustic loss increases by 20–40 dB/mm as the temperature is lowered from room temperature to $\sim$90 K. As the temperature is further reduced, the acoustic loss decreases from its peak value but remains higher than the room temperature value.

\begin{figure*}
\centering
\includegraphics[width=5.9in]{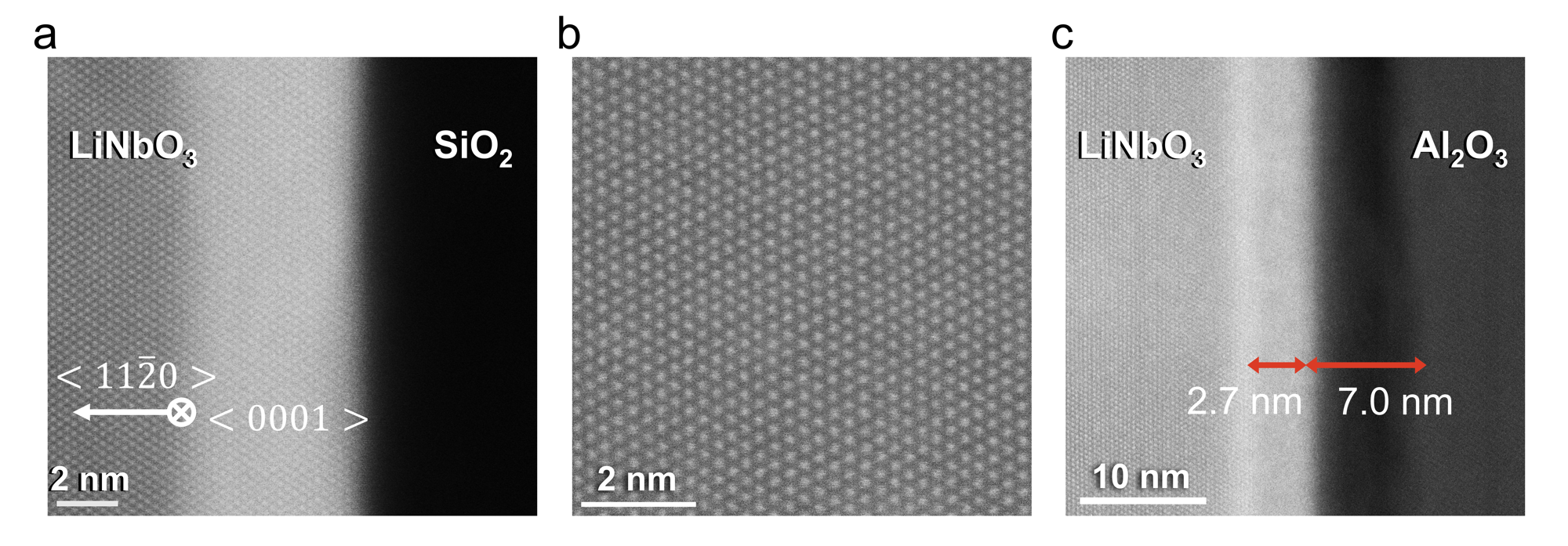}
\caption{Cross-sectional HAADF-STEM images of the TFLN sample along the $<0001>$ zone axis, showing (a) Lattice continuity at the bonding interface of LNOI; (b) Atomic lattice periodicity of lithium niobate on LNOI; (c) Lattice continuity at the bonding interface of LNOS.}
\label{Fig5}
\end{figure*}

We further calculate the quality factor $Q$ for the LNOI sample using the measured propagation loss (in dB/mm) shown in Fig. \ref{Fig3}(a). Here, the quality factor is defined as $Q=2\pi\,l/\lambda$, with $\lambda$ and $l$ denoting the acoustic wavelength and phonon mean free path, respectively. The quality factor is related to acoustic loss as $Q=\frac{0.0546\,f}{\alpha\,v_g}$, where $f$ is the acoustic frequency and $v_g$ is the group velocity. The extracted $Q$ factors are shown in Fig. \ref{Fig3}(b) and exhibit a temperature dependence similar to that reported for amorphous SiO\textsubscript{2} in previous work \cite{pohl2002low}. Interestingly, as the temperature is increased from room temperature to 350-400~K, the acoustic loss decreases. At 400~K, the measured loss ($4.56\pm0.33$~dB/mm) agrees with the frequency-dependent term shown in Fig.~\ref{Fig2}(d) within experimental uncertainty, suggesting that the frequency-independent loss channel may be suppressed at this temperature.

We also analyze the temperature-dependent acoustic propagation loss in bulk LN and suspended LN thin film. As shown in Fig.~\ref{Fig4}, the propagation losses measured on these two platforms show a linear temperature dependence from 4~K to room temperature, similar to LNOS. This behavior is consistent with Akhiezer damping. The acoustic propagation loss extrapolated to different frequencies is of the same order of magnitude as previously reported values \cite{slobodnik1970microwave,lu2020low}. The linear fit indicates that bulk and suspended LN exhibit similar temperature-dependent acoustic loss but differ slightly in the temperature-independent residual loss. The latter may arise from defect scattering in the crystal, scattering at the released interface, or strain variations induced during the release process.

Compared with bulk LN, the acoustic propagation losses in TFLN platforms are higher by several dB/mm. To further investigate the origin of this excess loss, we examine the crystal quality and bonding interfaces of the TFLN platforms using scanning transmission electron microscopy (STEM). The STEM imaging was performed on an aberration-corrected Nion UltraSTEM electron microscope operated at 60 kV, using a semi-convergence angle of 32 mrad. High-angle annular dark-field (HAADF) images were acquired with semi-collection angles ranging from 80-200 mrad, using a probe current of 10 pA. Fig. \ref{Fig5}(a), (b) shows STEM images of the bonding interface of the LNOI substrate, showing a good structural quality with a well-resolved LN lattice at the interface. An approximately 6 nm contrast variation is observed at the interface, which requires further analysis to determine its chemical composition. A similar bonding interface condition of LNOS is shown in Fig. \ref{Fig5}(c). Amorphization is observed at the interface, resulting in a brighter contrast of approximately 2.7~nm in the LN layer and a 7~nm-thick dark layer in the sapphire substrate. The darker contrast is likely due to ion milling during focused ion beam–transmission electron microscopy (FIB–TEM) sample preparation, which induces oxygen vacancies. 

Although the transition layer at the bonding interface is only a few nanometers thick, crystal impurities in this region can introduce elastic disorder and defect-induced scattering, thereby providing a measurable loss channel for SAWs. This mechanism may account for the observed temperature-independent propagation loss of approximately 1.97~dB/mm for the fundamental asymmetric Lamb (A0) mode in suspended LN and that of 1.11~dB/mm for the SH mode in LNOS. It may also contribute to the acoustic loss in LNOI.

In conclusion, we have systematically investigated acoustic loss mechanisms in thin-film lithium niobate (TFLN) platforms by comparing acoustic propagation loss across LNOI, LNOS, suspended LN thin films, and bulk LN over wide frequency and temperature ranges. We find that acoustic loss in LNOI exhibits a pronounced non-monotonic temperature dependence, which may arise from the amorphous BOX layer or the distinct multilayer structure compared to the other three platforms. At elevated temperatures, the acoustic loss in LNOI is suppressed. In contrast, acoustic loss in LNOS, bulk LN, and suspended LN thin films exhibits a monotonic decrease with decreasing temperature over the entire measured range, consistent with Akhiezer damping. The acoustic loss extracted from the free-propagating delay line represents an effective loss and may deviate from the intrinsic material loss due to acoustic walk-off and beam divergence. These effects can be mitigated using waveguide-based delay lines. However, this contribution does not affect the observed trend for each platform or the main conclusions of this work. 

In addition, high-resolution HAADF-STEM imaging reveals nanoscale interfacial disorder at the bonding interface, which may introduce an additional loss channel in TFLN devices relative to bulk LN. These results clarify the physical origins of acoustic loss in TFLN platforms and provide important material and structural guidelines for minimizing acoustic loss in future phononic, acousto-optic, piezo-optomechanics, and quantum acoustic devices.

\begin{acknowledgments}
This work is supported by the National Science Foundation (Award No. ITE-2134345 and OSI-2326746) and the DARPA MTO SOAR program (Award No. HR0011363032). Part of this work was conducted at the Washington Nanofabrication Facility and Molecular Analysis Facility, a National Nanotechnology Coordinated Infrastructure (NNCI) site at the University of Washington, with partial support from the National Science Foundation via award nos. NNCI-2025489. The electron microscopy part of this work was supported by the National Science Foundation (NSF) through a Materials Research Science and Engineering Center (DMR-2308979).
\end{acknowledgments}

\section*{Data availability}
The data that supports the findings of this study are available from the corresponding author upon reasonable request.

\bibliography{SAW_Loss_R1}

\clearpage
\onecolumngrid   

\thispagestyle{empty}
\vspace*{3cm}

\begin{center}
{\Large\bfseries Supplementary Information}\\[1cm]
for\\[0.5cm]
{\large\itshape Temperature-dependent acoustic loss at microwave frequencies in thin-film lithium niobate}\\[1cm]
Qixuan Lin \textit{et al.}
\end{center}

\clearpage

\setcounter{figure}{0}
\setcounter{table}{0}
\setcounter{equation}{0}
\setcounter{section}{0}

\renewcommand{\thefigure}{S\arabic{figure}}
\renewcommand{\thetable}{S\arabic{table}}
\renewcommand{\theequation}{S\arabic{equation}}
\renewcommand{\thesection}{S\arabic{section}}

\section{Device Fabrication}
\label{smsec:fab}
The devices on LNOI, LNOS, and bulk LN are patterned with electron-beam lithography (EBL) using a positive resist (ZEP520A). A 180-nm-thick aluminum layer is deposited by electron-beam evaporation, followed by lift-off in N-Methyl-2-pyrrolidone (NMP). 

\begin{figure*}[htbp]
\centering
\includegraphics[width=4in]{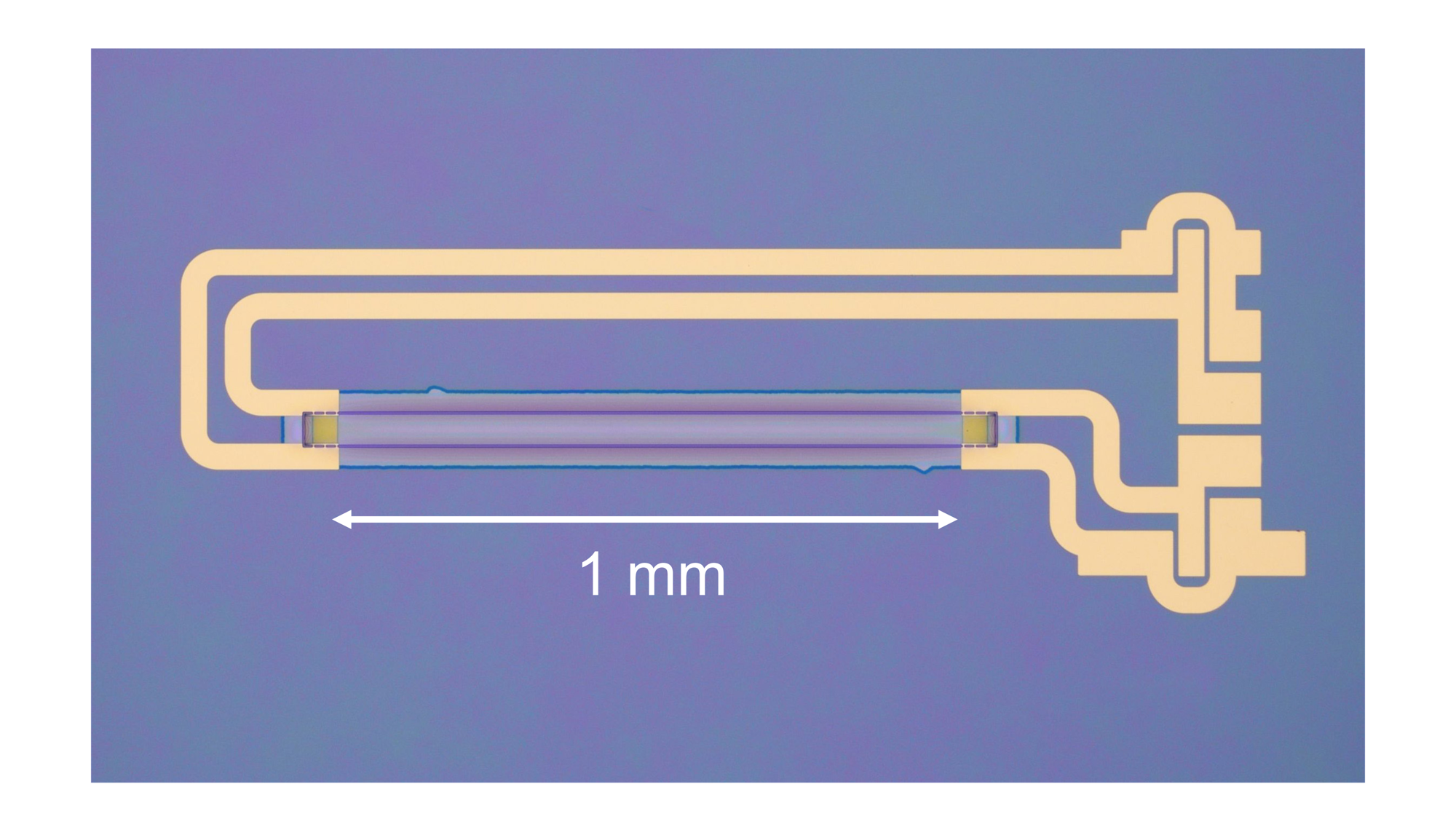}
\caption{\textbf{An optical image of a suspended LN device}.}
\label{FigS1}
\end{figure*}

The free-standing LN thin-film devices are fabricated on an LNOI substrate with a 300~nm-thick X-cut LN layer on a 2-~$\mu$m buried oxide on a silicon handle wafer. The release windows are patterned by EBL using ZEP520A, and then the LN layer is etched through by reactive-ion etching. After etching, the resist is stripped in NMP, and the chip is cleaned with piranha and standard cleaning solution to remove resist residue and redeposition. The IDT electrode was patterned with EBL and followed by a lift-off process of $5\ \mathrm{nm}$ Ti - $95\ \mathrm{nm}$ Au. The buried oxide layer was selectively under-etched in buffered oxide etchant (BOE) 10:1 to release the suspended LN membrane, followed by deionized water (DI) water rinses to remove debris. Finally, critical-point drying was performed to prevent membrane fracture due to surface-tension forces during liquid evaporation. An optical image of a suspended LN device is shown in Fig.~\ref{FigS1}.

\section{Calibration and Measurement Process}
\label{smsec:measurement}
The S-parameters of all devices from 4~K to 350~K were measured in a cryogenic probe station (Lakeshore CRX-4 K). The devices were probed using a pair of GSG probes, and each is connected to a channel of a VNA (Keysight N5230C PNA-L) for microwave reflection and transmission spectra measurements. Prior to the S-parameter measurements, the VNA was calibrated at room temperature to the probe tips using a calibration substrate (GGB Inc., CS-15). The calibration was verified at 4~K, showing no significant change in impedance over the measurement temperature range. No de-embedding process is used during data processing, as electrical parasitic are independent of the propagation length and therefore do not affect the extracted acoustic attenuation. The LNOI and bulk LN samples were measured at the same measurement run, while the LNOS and suspended LN samples were measured separately. The 400 K data of the LNOI Rayleigh mode is measured using a customized TEC-controlled heater.

\section{Effect of Acoustic Walk-Off and Beam Divergence on the Extracted Acoustic Loss}
\label{smsec:walkoff}
Acoustic walk-off, defined as the angular deviation between the phase-velocity direction determined by the input transducer and the direction of acoustic power flow, is an important consideration in LN acoustic devices. Here, we evaluate the walk-off angle $\phi$ using FEM simulations of the acoustic phase velocity for different phase velocity orientations $\theta$.

The walk-off angle can be analyzed using the slowness curve $s(\theta)=1/v_p(\theta)$, where $v_p(\theta)$ is the phase velocity of the acoustic wave. Because the acoustic power-flow density direction is determined by the group velocity direction, the walk-off angle can be obtained by computing the group velocity direction.

Assuming the acoustic wave propagates in-plane, with its phase propagating at an angle $\theta$ relative to the Y-axis for X-cut LN samples or relative to the X-axis for 128$^\circ$ YX-cut LN samples, the wavevector of the acoustic wave can be written as
\begin{equation}
    \label{eq:walkoff-k}
    \mathbf{k}(\theta)=(k_x,\,k_y)=\frac{\omega}{v_p(\theta)}(\cos\theta,\,\sin\theta)
\end{equation}

The group velocity can be calculated from $\mathbf{v_g}=\mathbf{\nabla_k}\omega$
\begin{equation}
    \label{eq:walkoff-vx}
    v_{g,x}(\theta)=\frac{\partial \omega}{\partial k_x}=\frac{\partial \omega}{\partial k}\frac{\partial k}{\partial k_x}+\frac{\partial \omega}{\partial \theta}\frac{\partial \theta}{\partial k_x}
    =v_p\cos\theta-v_p'\sin\theta
\end{equation}
\begin{equation}
    \label{eq:walkoff-vy}
    v_{g,y}(\theta)=\frac{\partial \omega}{\partial k_y}=v_p\sin\theta+v_p'\cos\theta
\end{equation}
where $v_p'=\partial v_p/\partial \theta$.

The walk-off angle $\phi$ can be calculate with the group velocity
\begin{equation}
    \label{eq:walkoff-angle_sum}
    \tan(\theta+\phi)=\frac{v_{g,y}}{v_{g,x}}=\frac{v_p\sin\theta+v_p'\cos\theta}{v_p\cos\theta-v_p'\sin\theta}
\end{equation}

For small walk-off, this simplifies to the commonly used approximation,
\begin{equation}
    \label{eq:walkoff-angle}
    \phi \approx \tan\phi = -\frac{s'(\theta)}{s(\theta)},
\end{equation}
where $s'=\partial s/\partial \theta$. Thus, the walk-off angle can be obtained directly from the angular derivative of the slowness curve or, equivalently, from the angular dependence of $v_p(\theta)$. 

\begin{figure*}[t]
\centering
\includegraphics[width=6in]{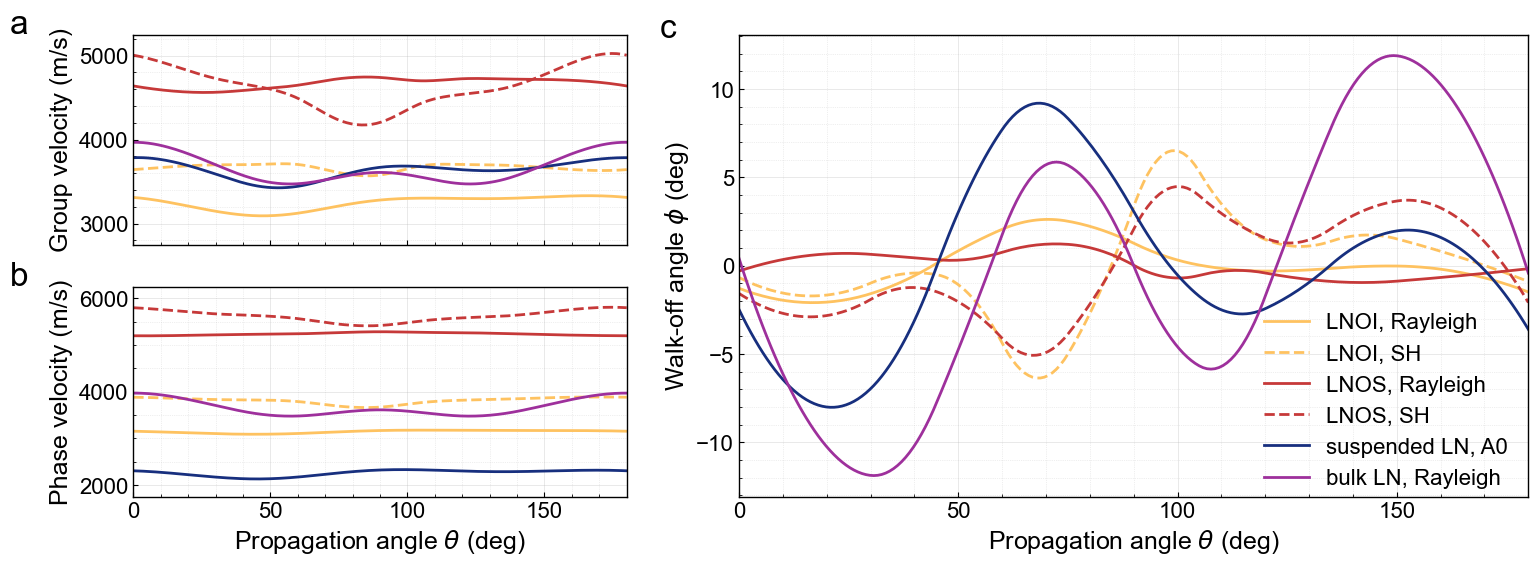}
\caption{\textbf{Acoustic velocity and walk-off angle simulations for different lithium-niobate platforms.} (a). Simulated acoustic group velocity as a function of the propagation angle. (b). Corresponding phase velocity versus propagation angle. (c). Walk-off angle extracted from the angular derivative of the slowness surface. The results includes LNOI (Rayleigh and SH at 1.8~GHz), LNOS (Rayleigh and SH at 1.8~GHz), bulk LN (Rayleigh at 2.1~GHz), and suspended LN (A0 at 1.8~GHz). For platforms with X-cut LN, including LNOI, LNOS, and suspended LN, the propagation angle $\theta$ is defined as the in-plane rotation relative to the Y crystal axis, whereas for the 128$^\circ$ YX-cut bulk LN, the angle is defined as the in-plane rotation relative to the X crystal axis.}
\label{FigS2}
\end{figure*}

We simulated the group velocities and phase velocities of various acoustic modes on different substrates as functions of the propagation angle $\theta$, as shown in Fig.~\ref{FigS2}(a) and (b). At $\theta=0^\circ$, which corresponds to the propagation direction used in the measurements, the simulated group velocities agree well with the experimentally extracted group velocities shown in Fig.~\ref{Fig2}(b), Fig.~\ref{FigS4}(c), Fig.~\ref{FigS5}(c), and Fig.~\ref{FigS6}(c). Using the phase-velocity results in Fig.~\ref{FigS2}(b) together with Eq.~\ref{eq:walkoff-angle}, we calculate the walk-off angles of the corresponding modes at different $\theta$, as shown in Fig.~\ref{FigS2}(c). 

At $\theta=0^\circ$, the Rayleigh mode in bulk 128$^\circ$YX-cut LN exhibits a near-zero walk-off angle. Although bulk X-cut LN can exhibit a larger walk-off when acoustic waves propagate along the Y-axis as reported\cite{slobodnik1970microwave}, the X-cut TFLN platforms studied here, including LNOI, LNOS, and suspended LN, show much smaller walk-off angles, ranging from $-0.3^\circ$ to $-2.5^\circ$.

In delay-line measurements, acoustic walk-off can introduce additional insertion loss that increases with the delay length due to the progressive misalignment between the acoustic energy flow and the receiving transducer aperture. The lateral shift of the acoustic beam caused by walk-off is approximately $\Delta x = L\tan\phi$. For small walk-off angles ($\phi \ll 1$), the overlap width between the acoustic beam and the receiving transducer aperture at a delay length $L$ can be approximated as

\begin{equation}
    w' \approx w - L\phi ,
\end{equation}

where $w$ is the aperture of the transducer and $w'$ is the effective overlap width between the acoustic beam and the receiving transducer aperture after propagating over a distance $L$. Taking into account the reduced receiving efficiency caused by acoustic walk-off, the transmission coefficient of the electrical signal through the first-arriving acoustic wave packet (Eq.~\ref{eq:first_arrive}) can be rewritten as

\begin{equation}
    t_{21}=\eta_1\eta_2 e^{-\alpha L}\left(1-\frac{\phi L}{w}\right)
    \approx
    \eta_1\eta_2 e^{-\alpha L-\phi L/w},
    \label{eq:first_arrive2}
\end{equation}

The second expression assumes that $\phi L/w$ is small, which is consistent with the experimental conditions.

The transducer aperture for the devices under test is designed to be $w=100\,\mu$m to mitigate acoustic beam divergence. For echoes reflected by the transducers, the walk-off angle remains the same for the backward propagation direction. As a result, the reflected acoustic waves propagate approximately along the same path, and the reflected beam remains fully captured by the transducer aperture after completing a round trip. Therefore, the transmission coefficient $t'_{21}$ in Eq.~\ref{eq:second_arrive} remains unchanged.

The ratio between the amplitudes of the two wave packets becomes

\begin{equation}
    \gamma(L)=e^{(\alpha-\phi/2w)2L}.
\end{equation}

We define the extracted effective propagation loss as $\alpha'$ and the intrinsic propagation loss due to intrinsic material properties as $\alpha$. The intrinsic propagation loss can therefore be obtained from the extracted value as

\begin{equation}
\alpha=\alpha'+\frac{\phi}{2w}.
\end{equation}

Using this expression together with the simulated walk-off angles, the intrinsic acoustic propagation loss of the corresponding modes presented in this work is theoretically underestimated by 0.49~dB/mm (LNOI, Rayleigh), 0.25~dB/mm (LNOI, SH), 0.60~dB/mm (LNOS, SH), 0.94~dB/mm (suspended LN, A0), and 0.16~dB/mm (bulk LN, Rayleigh) compared to the extracted value.

\begin{figure*}[htbp]
\centering
\includegraphics[width=4in]{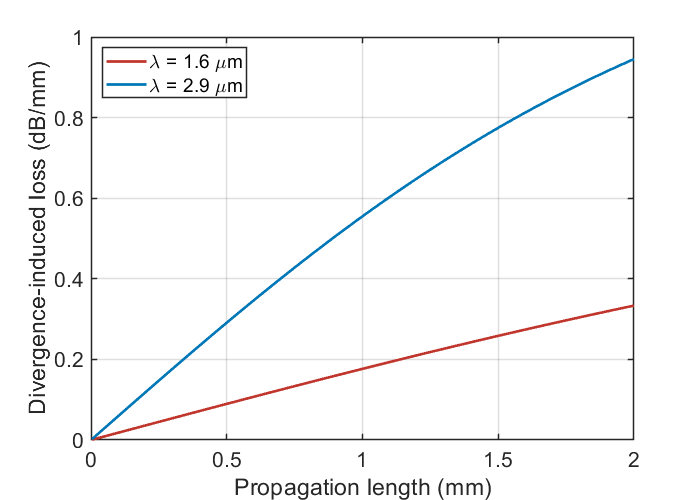}
\caption{\textbf{Calculated acoustic divergence-induced loss as a function of propagation length for acoustic wavelengths of 1.6~$\mu$m (blue) and 2.9~$\mu$m (red), assuming a device aperture of 100~$\mu$m.}}
\label{FigS3}
\end{figure*}

We use a simple Gaussian-overlap model to estimate the order of magnitude of the additional loss induced by beam divergence.
To estimate the additional loss induced by acoustic beam divergence in a free-propagating delay line, we approximate the acoustic beam using a Gaussian spreading model. The beam width evolves with propagation length \(L\) as
\begin{equation}
w(L)=w_0\sqrt{1+\left(\frac{4\,\lambda L}{\pi w_0^2}\right)^2},
\end{equation}
where \(w_0\) is the initial beam aperture, \(\lambda\) is the acoustic wavelength. 

Assuming that the receiving transducer has the same effective aperture as the initial beam width, the power overlap efficiency between the diverged acoustic beam and the receiving transducer is approximated as
\begin{equation}
\eta(L)=\left(\frac{w_0}{w(L)}\right)^2
=
\frac{1}{1+\left(\frac{4\,\lambda L}{\pi w_0^2}\right)^2}.
\end{equation}

The corresponding divergence-induced loss in dB/mm is then calculated as
\begin{equation}
\alpha_{\mathrm{div}}(L)=-10\log_{10}\eta(L)/L.
\end{equation}

We estimate the divergence-induced loss as a function of propagation length for acoustic wavelengths of 1.6~$\mu$m and 2.9~$\mu$m. As shown in Fig.~\ref{FigS3}, the divergence-induced loss is approximately 0.33~dB/mm and 0.94~dB/mm at a propagation length of 2~mm, which corresponds to the round-trip length of the devices measured in this work. These values represent an upper bound on the deviation of the extracted propagation loss due to beam divergence.

In conclusion, both acoustic walk-off and beam divergence contribute less than 1~dB/mm to the measured acoustic loss, with opposite effects. Since both effects are expected to have weak temperature dependence, the overall temperature-dependent trends and the conclusions for each platform are minimally affected.

\section{Representative Data for Each Sample}
\begin{figure*}[htbp]
\centering
\includegraphics[width=6.in]{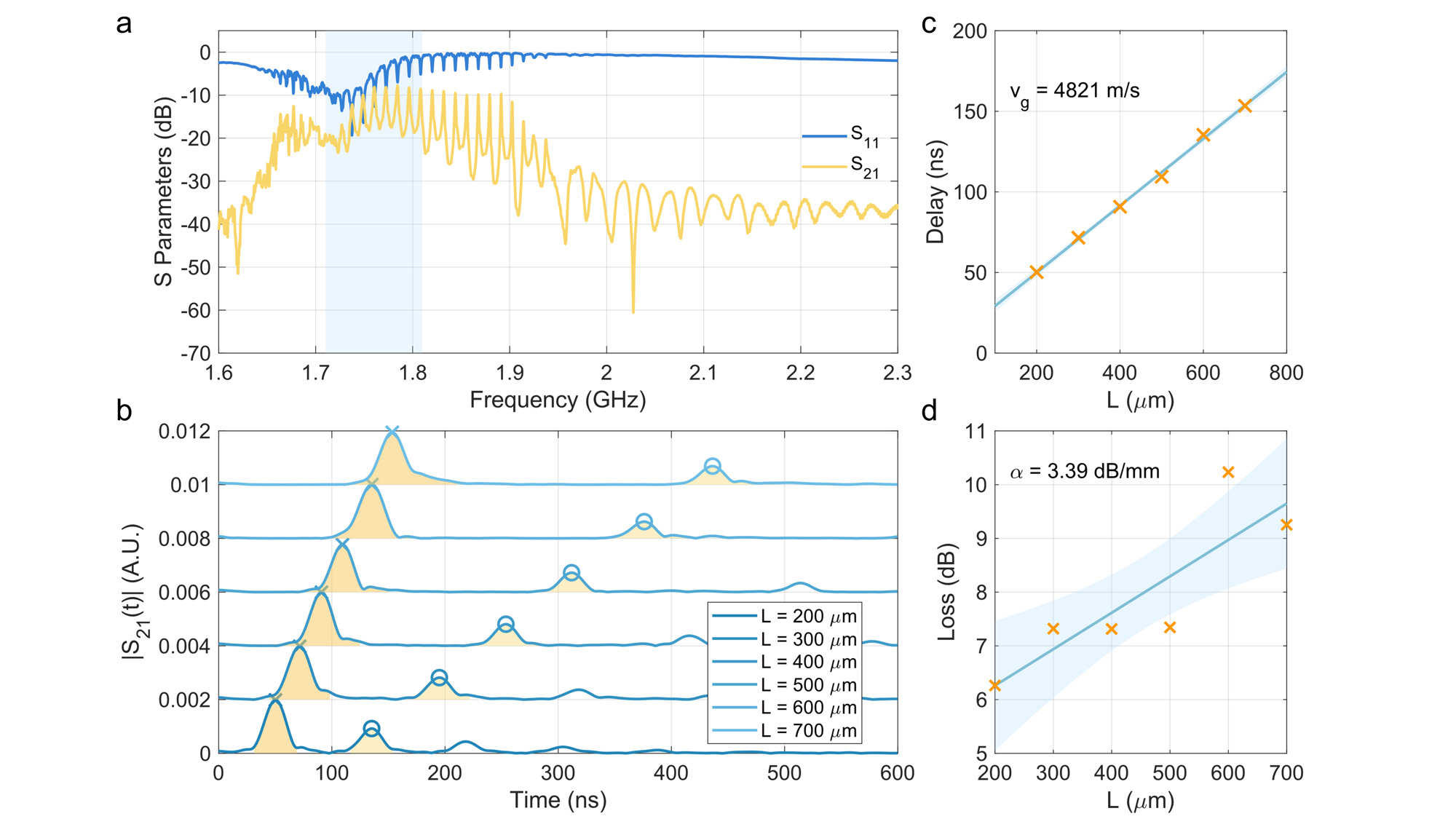}
\caption{\textbf{Representative data of LNOS devices.} (a). S-parameter measurement of the measured device with 200~$\mu\mathrm{m}$ delay length. The shaded region corresponds to the excitation of the SH acoustic mode. (b). Time-domain signals obtained by taking the inverse Fourier transform of the $S_{21}$ response within the shaded frequency range. Each trace corresponds to an IDT pair with a different delay length $L$, varying from $200~\mu \mathrm{m}$ (bottom) to $700~\mu \mathrm{m}$ (top), with vertical offsets applied for clarity. The shaded wave packets indicate acoustic pulses arriving at the receiving IDT via direct transmission or multiple round-trip reflections. Cross and circle markers denote the peak amplitudes of the first and the second arrival wave packets, respectively. Measurements are performed at room temperature. (c). Acoustic group velocity extracted from a linear fit of the delay time of the first wave packet as a function of the corresponding delay length. (d). The propagation loss extracted from a linear fit of the measured amplitude ratio between the first and the second wave packets at different delay lengths $L$.}
\label{FigS4}
\end{figure*}

\begin{figure*}[htbp]
\centering
\includegraphics[width=6.in]{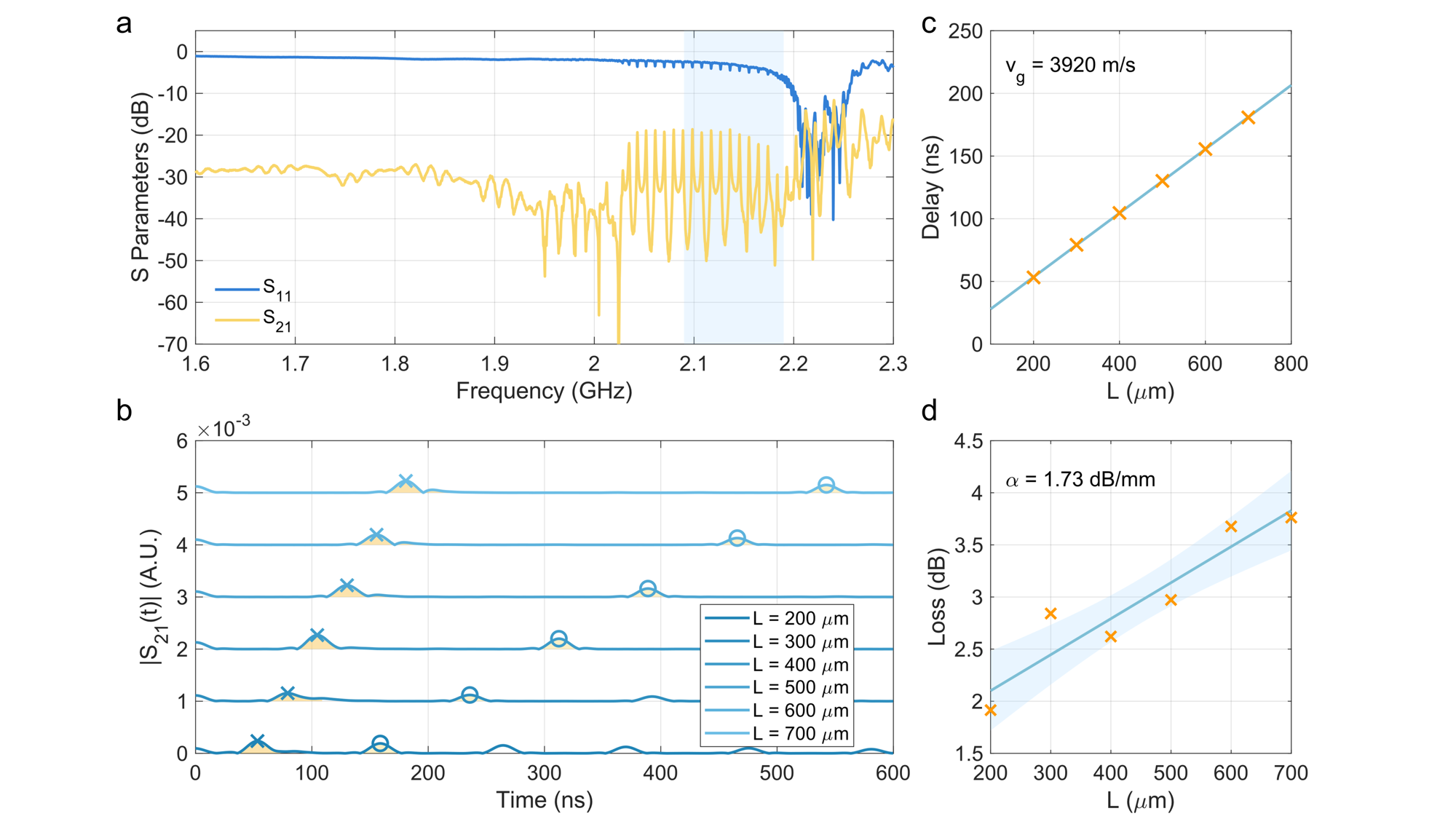}
\caption{\textbf{Representative data of bulk LN devices.} (a). S-parameter measurement of the measured device with 200~$\mu\mathrm{m}$ delay length. The shaded region corresponds to the excitation of the Rayleigh acoustic mode. (b). Time-domain signals obtained by taking the inverse Fourier transform of the $S_{21}$ response within the shaded frequency range. Each trace corresponds to an IDT pair with a different delay length $L$, varying from $200~\mu \mathrm{m}$ (bottom) to $700~\mu \mathrm{m}$ (top), with vertical offsets applied for clarity. The shaded wave packets indicate acoustic pulses arriving at the receiving IDT via direct transmission or multiple round-trip reflections. Cross and circle markers denote the peak amplitudes of the first and the second arrival wave packets, respectively. Measurements are performed at room temperature. (c). Acoustic group velocity extracted from a linear fit of the delay time of the first wave packet as a function of the corresponding delay length. (d). The propagation loss extracted from a linear fit of the measured amplitude ratio between the first and the second wave packets at different delay lengths $L$.}
\label{FigS5}
\end{figure*}

\begin{figure*}[htbp]
\centering
\includegraphics[width=6.in]{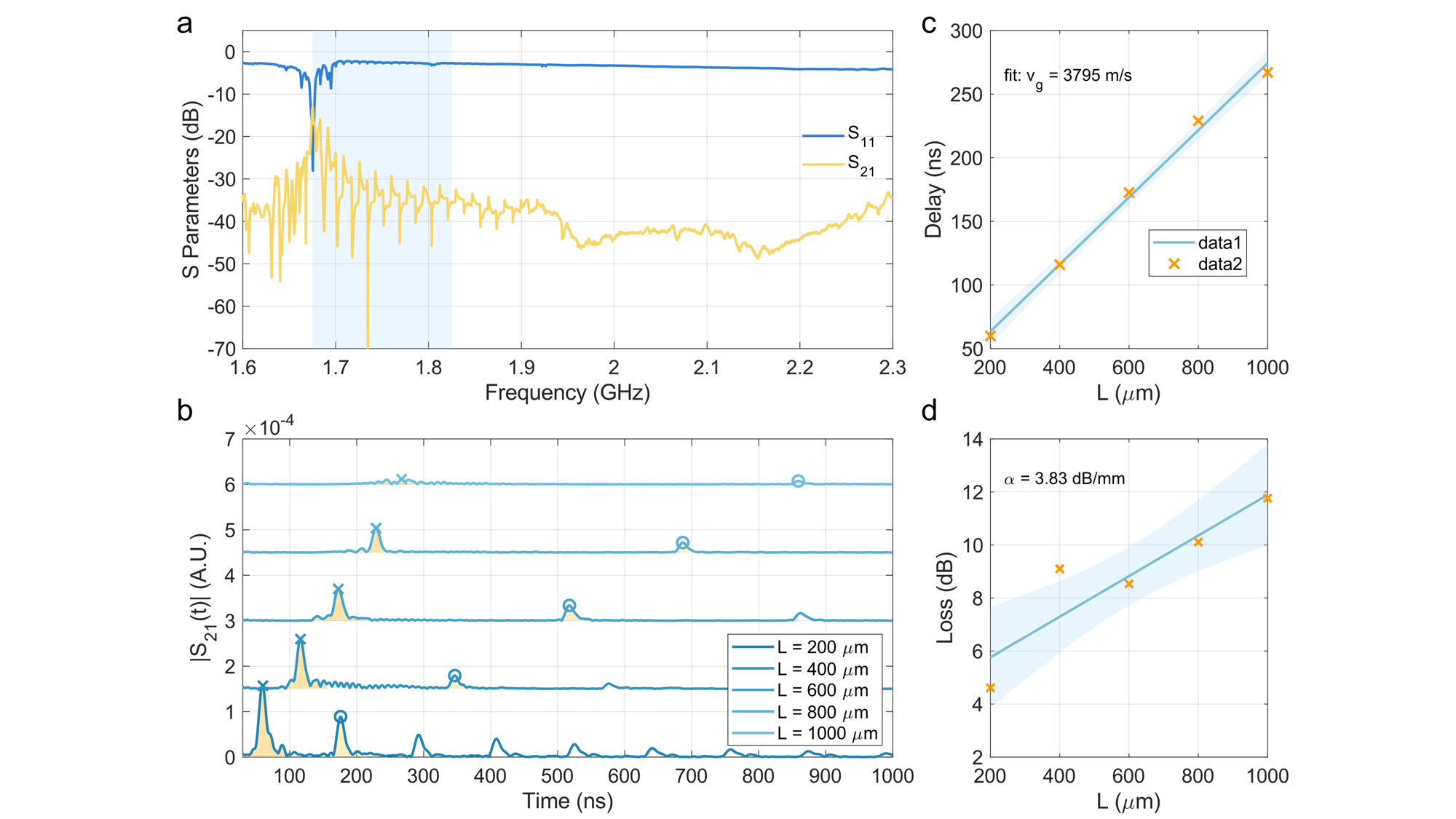}
\caption{\textbf{Representative data of suspended LN devices.} (a). S-parameter measurement of the measured device with 200~$\mu\mathrm{m}$ delay length. The shaded region corresponds to the excitation of the A0 acoustic mode. (b). Time-domain signals obtained by taking the inverse Fourier transform of the $S_{21}$ response within the shaded frequency range. Each trace corresponds to an IDT pair with a different delay length $L$, varying from $200~\mu \mathrm{m}$ (bottom) to $1000~\mu \mathrm{m}$ (top), with vertical offsets applied for clarity. The shaded wave packets indicate acoustic pulses arriving at the receiving IDT via direct transmission or multiple round-trip reflections. Cross and circle markers denote the peak amplitudes of the first and the second arrival wave packets, respectively. Measurements are performed at room temperature. (c). Acoustic group velocity extracted from a linear fit of the delay time of the first wave packet as a function of the corresponding delay length. (d). The propagation loss extracted from a linear fit of the measured amplitude ratio between the first and the second wave packets at different delay lengths $L$.}
\label{FigS6}
\end{figure*}

We present representative data, including S-parameters, time-domain signals of  $S_{21}$, extracted group velocities, and acoustic loss at the corresponding frequencies for the SH mode on LNOS (Fig.~\ref{FigS4}), the Rayleigh mode on bulk LN (Fig.~\ref{FigS5}), and the A0 mode on suspended LN (Fig.~\ref{FigS6}) at room temperature.

In Fig.~\ref{FigS4}(a), Fig.~\ref{FigS5}(a), and Fig.~\ref{FigS6}(a), the dips in the S-parameters arise from the resonant effect of the cavity formed by the IDT pair. The free spectral range (FSR) scales with the delay-line length. The acoustic mode in the analyzed frequency region (shaded region) is identified based on the extracted group velocities shown in Fig.~\ref{FigS4}(c), Fig.~\ref{FigS5}(c), and Fig.~\ref{FigS6}(c), respectively.

The time-domain signals obtained by taking the inverse Fourier transform of the $S_{21}$ response within the shaded frequency range are shown in Fig.~\ref{FigS4}(b), Fig.~\ref{FigS5}(b), and Fig.~\ref{FigS6}(b). Acoustic loss as a function of delay length and the fitted propagation loss of the selected mode of each platform is shown in  Fig.~\ref{FigS4}(d), Fig.~\ref{FigS5}(d), and Fig.~\ref{FigS6}(d).

\end{document}